\documentclass[a4paper,11pt]{article}
\pdfoutput=1 

\usepackage{jheppub} 

\usepackage[T1]{fontenc} 
\usepackage{cleveref}

\newcommand{\la}{\langle}
\newcommand{\ra}{\rangle}
\newcommand{\Nt}[2]{\tilde{\nabla}_{#1,#2}}

\title{Soft Theorems and the KLT-Relation}

\author[a]{Rafael Aoude}
\author[b]{and Andreas Helset}

\affiliation[a]{PRISMA$^+$ Cluster of Excellence \& Institute of Physics, \\ Johannes Gutenberg-Universit\"{a}t Mainz, 55099 Mainz, Germany} 
\affiliation[b]{Niels Bohr International Academy and Discovery Center, Niels Bohr Institute, University of Copenhagen, Blegdamsvej 17, DK-2100 Copenhagen, Denmark}

\emailAdd{aoude@uni-mainz.de}
\emailAdd{ahelset@nbi.ku.dk}
\preprint{MITP/19-077}

\abstract{
We find new relations for the non-universal part of the Yang-Mills amplitudes by combining the 
KLT-relation and the soft behavior of gauge and gravity amplitudes.
We also extend the relations to include contributions from effective operators.
}

\begin{document} 
\maketitle
\flushbottom

\section{Introduction}
\label{sec:intro}

The study of scattering amplitudes when the momentum of one or more particles 
becomes soft has a long history \cite{Bloch:1937pw,Nordsieck:1937zz,Low:1954kd,GellMann:1954kc,Low:1958sn,Yennie:1961ad,Kinoshita:1962ur,Lee:1964is,Weinberg:1964ew,Weinberg:1965nx,Burnett:1967km,Bell:1969yw}. Weinberg showed that the scattering amplitudes
factorize when a photon or graviton becomes soft, and that this factorization is universal \cite{Weinberg:1964ew,Weinberg:1965nx}. 
The universality of the soft photon theorem is due to charge conservation, while for a soft graviton
it follows from the equivalence principle.
A subleading soft theorem for photons at tree-level was proven by Low \cite{Low:1958sn}.
Similar subleading soft theorems for gravitons and gluons have more recently been discussed using eikonal methods \cite{Laenen:2008gt,Laenen:2010uz,White:2011yy}.
Also, Cachazo and Strominger showed that the sub-subleading soft graviton correction at tree-level is also universal \cite{Cachazo:2014fwa}.

The soft graviton theorem was shown to be connected to the Bondi, van der Burg, Metzner 
and Sachs (BMS) symmetry \cite{Bondi:1962px,Sachs:1962wk}, as a Ward identity \cite{Strominger:2013jfa,He:2014laa,Kapec:2014opa}.
This has sparked an interest in the connection between asymptotic symmetries and soft theorems (see Ref.~\cite{Strominger:2017zoo} for a list of references). The subleading soft theorem is known to be related to the supertranslations
and superrotations for asymptotic symmetries and the sub-subleading soft theorem related symmetries was recently analysed in Ref.~\cite{Campiglia:2016efb,Campiglia:2016jdj}. The authors found a new class of vector fields, which 
hints in the direction of a BMS algebra extension.

At loop level, the leading soft theorems for photons and gravitons remain unchanged. However, the
loop corrections to the subleading soft theorems for gluons and gravitons were discussed in Refs.~\cite{Bern:2014oka,He:2014bga}.

The soft behavior of scattering amplitudes when more than one particle is taken soft has 
also been studied (see e.g. Ref.~\cite{AtulBhatkar:2018kfi} and references therein).
The soft theorems have also been discussed outside four dimensions \cite{Schwab:2014xua,Afkhami-Jeddi:2014fia,Zlotnikov:2014sva,Kalousios:2014uva} using the scattering equation framework by Cachazo, He, and Yuan (CHY) \cite{Cachazo:2013gna,Cachazo:2013hca,Cachazo:2013iea}.

Much of the recent progress in calculating gravitational scattering amplitudes relies the connection between
gauge amplitudes and gravity amplitudes \cite{Bern:2010ue,Bern:2019prr}. One manifestation of this connection is the Kawai-Lewellen-Tye (KLT) 
relations \cite{Kawai:1985xq}, which relates open and closed string amplitudes at tree level. 
In the field theory limit, the KLT-relation states that a sum of products of two color-ordered 
Yang-Mills amplitudes is a gravity amplitude.

In this paper, we study the connection between gravity and gauge soft theorems via the KLT-relations.
	We compare both sides of the formula at sub-subleading order in the soft-momentum expansion and obtain relations for the \textit{non-universal} part of the Yang-Mills amplitude.
To the best of our knowledge, no other relations have been obtained for the non-universal piece of the Yang-Mills amplitudes. We further
study the insertion of effective operators, which start contributing at sub-leading order. We also obtain relations for the
non-universal effective amplitude at sub- and sub-subleading order.

The paper is organized as follows: \Cref{sec:BCFW} reviews the derivation of the soft theorems using 
Britto-Cachazo-Feng-Witten (BCFW) 
recursion relations. \Cref{sec:SoftTheorem} presents the soft theorems of Yang-Mills and gravity amplitudes,
while \cref{sec:KLTrelation} introduces the KLT-relations. Two of the new results of the paper are presented
in \cref{sec:Subsubleading}, where the soft limit of the amplitudes and the KLT-relation are used to
find non-trivial relations which the Yang-Mills amplitudes must satisfy.
An extension of these results is presented in \cref{sec:EffectiveOperators}, when effective operators are
included. We conclude in \cref{sec:Conclusion}.


\section{BFCW}
\label{sec:BCFW}

We will review the derivation of the soft theorems for Yang-Mills and gravity amplitudes using the BCFW recursion
relations \cite{Britto:2004ap,Britto:2005fq}. This follows closely the derivation of the new soft theorems by Cachazo and Strominger \cite{Cachazo:2014fwa}. 
We use the spinor-helicity formalism, with the convension $s_{ab} = \langle a,b\rangle[b,a]$.
For an $(n+1)$-point amplitude with a soft particle $s$ with positive helicity,\footnote{The 
negative helicity case can be derived analogously.} $\mathcal{A}_{n+1}(s,1,\dots,n)$, we perform
the BCFW shift 
\begin{align}
	\lambda_s(z) = \lambda_s + z\lambda_n, \qquad \tilde\lambda_n(z) 
	= \tilde \lambda_n - z \tilde\lambda_s.
\end{align}
The original amplitude can be recovered from the complex deformed amplitude as the residue at $z=0$,
\begin{align}
	\mathcal{A}_{n+1} = \frac{1}{2\pi i}\int \frac{dz}{z} \mathcal{A}_{n+1}(z).
\end{align}
Using Cauchy's residue theorem, we find the following relation
\begin{align}
	\label{eq:BCFWsum}
	\mathcal{A}_{n+1} = \sum_{\textrm{diagrams }I}\mathcal{A}_L(z_I) \frac{1}{P_I^2}\mathcal{A}_R(z_I),
\end{align}
where the amplitude factorizes into a sum of two lower-point amplitudes, assuming the contribution from $|z|\rightarrow \infty$ vanishes. 
The product of the two amplitudes is the residue in Cauchy's theorem,
therefore they are evaluated at the pole $z_I$. The pole is found 
for each diagram by solving the equation $P^2_{I}(z_I) = 0$. From this relation we can 
build up higher-point amplitudes recursively.

The sum in \cref{eq:BCFWsum} can be split into two parts as
\begin{align}
	\label{eq:BCFWsplit}
	\mathcal{A}_{n+1} = \mathcal{A}_3(s(z_I), 1, I)\frac{1}{P_I^2}\mathcal{A}_n(-I,2,\dots,n(z_I)) + \mathcal{R}_{n+1}.
\end{align}
%
The remainder term is written as
\begin{align}
	\label{eq:Rdef}
	\mathcal{R}_{n+1} = \sum_{\textrm{diagrams I}} \mathcal{A}_L(s(z_I),1,2,\dots,j)\frac{1}{P_I^2}
	\mathcal{A}_R(j+1,\dots,n(z_I))
\end{align}
where $\mathcal{A}_L$ in \cref{eq:Rdef} is a four- or higher-point amplitude. Under the holomorphic scaling $\lambda_s\rightarrow\epsilon
\lambda_s$, $\tilde\lambda_s\rightarrow\tilde\lambda_s$, the amplitude\footnote{We will denote a Yang-Mills amplitude by $\mathcal{A}$, and a gravity amplitude by $\mathcal{M}$.} for a soft gluon takes the form
\begin{align}
	\mathcal{A}_{n+1}&= 	\frac{1}{\epsilon^2}\frac{\langle n1\rangle}{\langle ns\rangle\langle s1\rangle}
	\mathcal{A}_n\left(\{\lambda_1,\tilde\lambda_1 + \epsilon\frac{\langle ns\rangle}{\langle n1\rangle}
\tilde\lambda_s\}, \dots,\{\lambda_n,\tilde\lambda_s + \epsilon\frac{\langle s1\rangle}
{\langle n1\rangle}\tilde\lambda_s\}\right) +\mathcal{R}_{n+1} \nonumber \\
	&= \frac{1}{\epsilon^2}\frac{\langle n1\rangle}{\langle ns\rangle\langle s1\rangle}
	e^{\epsilon\left(\frac{\langle ns\rangle}{\langle n1\rangle}\tilde\lambda_s
	\frac{d}{d\tilde\lambda_1} + \frac{\langle s1\rangle}{\langle n1\rangle}
	\tilde\lambda_s\frac{d}{d\tilde\lambda_n}\right)}
\mathcal{A}_n +  \mathcal{R}_{n+1}.
\end{align}
This form was written down in Ref.~\cite{Du:2014eca}. Also, the authors of Ref.~\cite{Cachazo:2014fwa} showed that $\mathcal{R}_{n+1}$ is of order 
$\mathcal{O}(\epsilon^0)$.
A similar expression can be written down for gravity, with the leading pole being 
$\mathcal{O}(\epsilon^{-3})$. By expanding the exponential we can find the universal 
leading and (at tree-level) subleading soft factors, $S^{(0)}_{\textrm{YM}}$ and $S^{(1)}_{\textrm{YM}}$. The exponential 
contains derivative terms, for which we use the notation $\Nt{a}{b} = \tilde{\lambda}^{\dot \alpha}_{a}\frac{d}{d\tilde{\lambda}^{\dot\alpha}_b}$.
The soft factors can be written as
\begin{align}
	\label{eq:SoftFactorYM}
	S^{(k)}_{\textrm{YM}} = \frac{\langle n1\rangle}{\langle ns\rangle\langle s1\rangle}
	\frac{1}{k!}\left(\frac{\langle ns\rangle}{\langle n1\rangle}\Nt{s}{1} + \frac{\langle s1\rangle}{\langle n1\rangle}
	\Nt{s}{n}\right)^k,
\end{align}
where the soft factors with $k\geq 2$ give only a part of the amplitude.
Again, a similar partial infinite soft factor for gravity can be found analogously.

A partial infinite soft theorem for effective operators can also be found using the same method.
For Yang-Mills, we find that
%
\begin{align}
\label{eq:SoftTermEff}
	\overline S^{(k+1)}_{\textrm{YM}} = -\left(\frac{[ns]}{\langle ns\rangle}
	+\frac{[1s]}{\langle 1s\rangle}\right)
	\frac{1}{k!}\left(\frac{\langle ns\rangle}{\langle n1\rangle}\Nt{s}{1} + \frac{\langle s1\rangle}{\langle n1\rangle}
	\Nt{s}{n}\right)^k.
\end{align}
The $k=0$ term reproduces the leading soft theorem for effective operators with a spin-1 particle discussed in Ref.~\cite{Elvang:2016qvq}. 
The first contribution from effective operators for a Yang-Mills amplitude appears at subleading order. 
For gravity, the first contribution from effective operators only enters at sub-subleading order.


\section{Soft Theorem}
\label{sec:SoftTheorem}

We will show the connection between the soft factors of gauge theory and gravity using the 
KLT-formula. In particular, at sub-subleading order in the soft expansion
we find new relations between tree-level amplitudes. We start with 
an amplitude with one external soft particle. 
The soft limits of an amplitude with an external soft spin-1/spin-2 particle are 
\begin{align}
	\mathcal{A}_{n+1} &= \left(\frac{S_{\text{YM}}^{(0)}}{\epsilon^2}
	+ \frac{S_{\text{YM}}^{(1)}}{\epsilon} + S_{\textrm{YM}}^{(2)}\right)\mathcal{A}_{n} +\mathcal{R}_{n+1} + \mathcal{O}({\epsilon^1}),\\
	\mathcal{M}_{n+1} &= \left(\frac{S_{\text{GR}}^{(0)}}{\epsilon^3}
+ \frac{S_{\text{GR}}^{(1)}}{\epsilon^2} 
+ \frac{S_{\text{GR}}^{(2)}}{\epsilon}\right)\mathcal{M}_{n} + \mathcal{O}({\epsilon^0}),
\end{align}
respectively.
The $S_{\text{YM}}^{(i)}$ is the $i$th subleading soft factor of an amplitude with a soft spin-1 particle, and similarly for gravity.
The \textit{non-universal} part of the Yang-Mills amplitude enters at sub-subleading order.
In the KLT-formula, two Yang-Mills amplitudes with different color-ordering are required: $\mathcal{A}_{n}(t,\sigma,n-1,n)$ and
 $\mathcal{\tilde{A}}_{n}(n-1,\rho,n,t)$.
 From now on we assume that whenever $\mathcal{A}_n$ and $\mathcal{\tilde{A}}_{n}$ is written, we have this particular ordering. 
 The same holds for the $(n+1)$-point amplitudes; the relevant color-orderings are $\mathcal{A}_{n+1}(t,\sigma,n-1,n,n+1)$, $\mathcal{\tilde{A}}_{n+1}(n-1,\rho,n,t,n+1)$,
$\mathcal{R}_{n+1}(t,\sigma,n-1,n,n+1)$ and $\mathcal{\tilde{R}}_{n+1}(n-1,\rho,n,t,n+1)$. We leave the ordering
implicit from now on.

%
The soft limit of the Yang-Mills amplitude is given by the soft factors in \cref{eq:SoftFactorYM}.
The soft limit of the gravity amplitude is
%
\begin{align}
	\label{eq:SoftGravityAmplitude}
	\mathcal{M}&_{n+1}(1,2,\dots,n+1) =- 
	\frac{1}{\epsilon^3}\sum_{k=1}^{n}\frac{[n+1,k]}{\langle n+1,k \rangle}\\
	&\left[\frac{\langle x,k\rangle\langle y,k\rangle}{\langle x,n+1 \rangle \langle y,n+1\rangle} 
		\frac{\epsilon}{2}\left(\frac{\langle x,k \rangle}{\langle x,n+1 \rangle}
		+ \frac{\langle y,k \rangle}{\langle y,n+1 \rangle} \right) \Nt{n+1}{k}
		+ \frac{\epsilon^2}{2}\left(\Nt{n+1}{k}\right)^2
	\right]\mathcal{M}_n(1,2,\dots,n), \nonumber 
\end{align}
where $x,y$ are reference spinors which specifies a gauge. The amplitudes are independent of the choice of $x,y$.
Some care is needed when implementing the momentum conservation. We use $(n+1)$-point and $n$-point momentum
conservation, given by
\begin{align}
	\label{eq:MomConsN+1}
	\tilde\lambda_{i} = -\sum_{\underset{k\neq i,j}{k=1}}^{n}\frac{\langle j,k\rangle}{\langle j,i\rangle} \tilde\lambda_k
	- \epsilon \frac{\langle j,n+1 \rangle}{\langle j,i\rangle}\tilde\lambda_{n+1}, \qquad
	\tilde\lambda_{j} = -\sum_{\underset{k\neq i,j}{k=1}}^{n}\frac{\langle i,k\rangle}{\langle i,j\rangle} \tilde\lambda_k
	- \epsilon \frac{\langle i,n+1 \rangle}{\langle i,j\rangle}\tilde\lambda_{n+1}.
\end{align}
and
\begin{align}
	\label{eq:MomConsN}
	\tilde\lambda_{i} = -\sum_{\underset{k\neq i,j}{k=1}}^{n}\frac{\langle j,k\rangle}{\langle j,i\rangle} \tilde\lambda_k, \qquad
	\tilde\lambda_{j} = -\sum_{\underset{k\neq i,j}{k=1}}^{n}\frac{\langle i,k\rangle}{\langle i,j\rangle} \tilde\lambda_k.
\end{align}
We use eq.~\eqref{eq:MomConsN+1} for all $(n+1)$-point amplitudes $\mathcal{A}_{n+1}$ and 
$\mathcal{M}_{n+1}$,  and 
eq.~\eqref{eq:MomConsN} for all $n$-point amplitudes $\mathcal{A}_n$ and $\mathcal{M}_n$, with $i,j=n-1,n$.
Also, the total derivatives in the soft factors are
\begin{align}
	\Nt{s}{k}=\tilde\lambda_s^{\dot\alpha}\frac{d}{d\tilde\lambda^{\dot\alpha}_k} =\tilde\lambda_s^{\dot\alpha}\left[ \frac{\partial}{\partial \tilde\lambda^{\dot\alpha}_k} 
		+ \left(-\frac{\langle j,k\rangle}{\langle j,i\rangle}\right) \frac{\partial}{\partial\tilde\lambda^{\dot\alpha}_i}
	+ \left(-\frac{\langle i,k\rangle }{\langle i,j\rangle}\right)\frac{\partial}{\partial \tilde\lambda^{\dot\alpha}_j}\right].
\end{align}
When first using momentum conservation before applying the soft factors, the total derivatives reduce to partial
derivatives.


\section{KLT-Relation}
\label{sec:KLTrelation}

In string theory, the KLT-relation provides a connection between open and closed string amplitudes. In the limit
of infinite string tension, field theory is recoved and a relation between gravity and gauge amplitudes is obtained. Once
all the proper permutations are taken into account, the KLT-relation gives the gravity amplitude as the "square" of
the gauge amplitudes. For low-point amplitudes, the formulas are relatively simple, which helps streamlining gravitational scattering-amplitude calculations.

The most general form of the KLT-relation is \cite{Bern:1998sv,BjerrumBohr:2010yc}
\begin{align}
\label{eq:KLTgeneral}
 \mathcal{M}_n(1,2,\dots,n) = &(-1)^{n+1} \sum_{\sigma\in S_{n-3}}\sum_{\alpha\in S_{j-1}}\sum_{\beta\in S_{n-2-j}}
 \mathcal{A}_n(1,\sigma_{2,j},\sigma_{j+1,n-2},n-1,n) 
 \\
 \times \mathcal{S}[\alpha_{\sigma(2),\sigma(j)}|\sigma_{2,j}]_{p_{1}}
 &\mathcal{S}[\sigma_{j+1,n-2}|\beta_{\sigma(j+1),\sigma(n-2)}]_{p_{n-1}}                                        
 \mathcal{\tilde{A}}_n(\alpha_{\sigma(2),\sigma(j)},1,n-1,\beta_{\sigma(j+1),\sigma(n-2)},n), \nonumber 
\end{align}
where $\alpha,\beta,\sigma, \rho$ are particular orderings of the color-ordered Yang-Mills amplitudes.
The KLT-kernel $\mathcal{S}$ is defined as
\begin{align}
	\label{eq:kernelDef}
	\mathcal{S}[i_1,\dots,i_k|j_1,\dots,j_k]_{p_1} = \prod_{t=1}^{k} ( s_{i_t1} + \sum_{q>t}^k \theta(i_t,i_q)s_{i_ti_q}),
\end{align}
where $\theta(i_a,i_b)$ is $0$ if $i_a$ sequentially comes before $i_b$ in the set $\{j_1,\dots,j_k\}$,
and otherwise it takes the value $1$. One of the properties of the kernel is to take into account
the fact that Yang-Mills amplitudes are color-order while gravity amplitudes are not. It was also proven in
Ref.~\cite{BjerrumBohr:2010yc} that the KLT-relation, as written in
eq.~\eqref{eq:KLTgeneral}, is independent of the choice of $j$. Therefore, with $j=2$, we have for an $(n+1)$-point amplitude that 
\begin{align}
        \label{eq:KLTj2}
	\mathcal{M}_{n+1}(1,2,\dots,n,n+1)= (-1)^{n} \sum_{t=1}^{n-2} \sum_{\sigma,\rho\in S_{n-3}}
	&\mathcal{A}_{n+1}(t,\sigma,n-1,n,n+1) \mathcal{S}[t|t]_{p_{n+1}} \mathcal{S}[\sigma|\rho]_{p_{n-1}}
	\nonumber \\
	&\times \mathcal{\tilde{A}}_{n+1}(n-1,\rho,n,t,n+1).
\end{align}
In the next section, we are going to apply the soft theorems for each amplitude and collect
terms at different orders in $1/\epsilon$. Thus, we also need the soft limit of the KLT-kernel, which is \cite{Du:2014eca}
\begin{align}
	\label{eq:SoftKernel}
	\mathcal{S}[t|t]_{p_{n+1}}\mathcal{S}[\sigma|\rho]_{p_{n-1}} \rightarrow 
	\epsilon s_{t,n+1} e^{\epsilon \frac{\langle n,n+1\rangle}{\langle n,t\rangle} 
	\Nt{n+1}{t}}\mathcal{S}[\sigma|\rho]_{p_{n-1}}.
\end{align}
We also have the $S_{n-3}-$symmetric form of the KLT-relation for $n$-point amplitudes
\begin{align}
        \label{eq:KLTsymmetric}
	\mathcal{M}_n(1,2,\dots,n)= (-1)^{n+1} \sum_{\sigma,\rho\in S_{n-3}} \mathcal{A}_n (1,\sigma,n-1,n)
	\mathcal{S}[\sigma|\rho]_{p_{n-1}} \mathcal{\tilde{A}}_n(1,n-1,\rho,n).
\end{align}
The different forms of the KLT-relation will be useful shortly.


\section{Non-universal Relations}
\label{sec:Subsubleading}
The usual procedure when using the KLT-relation is to obtain gravity amplitudes from Yang-Mills amplitudes, 
since usually the Yang-Mills amplitudes are easier to calculate.
 Here, we go in the opposite direction. We use information about the gravity amplitudes to
obtain relations on the Yang-Mills side. As we noted before, the \textit{non-universal} part of the 
Yang-Mills amplitude enters
at sub-subleading order in the soft-momentum expansion. At this order, we also have a universal part which 
comes from an exponential of the associated soft factor. 
Both terms contribute in the KLT-formula. 
On the other hand, gravity contains only \textit{universal} pieces at $\mathcal{O}(1/\epsilon)$ . 
We equate the soft limit of the gravity amplitude with the soft limit of the Yang-Mills side in the KLT-relation.
This immediately gives constraints for the \textit{non-universal} part of the Yang-Mills amplitudes. 
We describe the procedure in the following and give a detailed derivation in \Cref{app:subsubleading}.

We use the KLT-relation in \cref{eq:KLTj2}, which we write as
\begin{align}
	\label{eq:leftsiderightside}
	\mathcal{M}_{n+1} = \sum_{\sigma,\rho}\mathcal{A}_{n+1}(\sigma) \mathcal{S}_{n+1}[\sigma|\rho]
	\mathcal{\tilde{A}}_{n+1}(\rho).
\end{align}
For the left-hand side of \cref{eq:leftsiderightside}, we apply the soft-graviton theorem in \cref{eq:SoftGravityAmplitude}, and then apply 
the KLT-relation in \cref{eq:KLTsymmetric} for each $k$ in the sum for the soft factor,
\begin{align}
	\mathcal{M}_{n+1} &= \left(\frac{S_{\textrm{GR}}^{(0)}}{\epsilon^3} + \frac{S_{\textrm{GR}}^{(1)}}{\epsilon^2} 
	+\frac{S_{\textrm{GR}}^{(2)}}{\epsilon}\right)\mathcal{M}_n \nonumber \\
	&= \left(\frac{S_{\textrm{GR}}^{(0)}}{\epsilon^3} + \frac{S_{\textrm{GR}}^{(1)}}{\epsilon^2} 
	+\frac{S_{\textrm{GR}}^{(2)}}{\epsilon}\right) \sum_{\alpha,\beta}\mathcal{A}_n(\alpha)
	\mathcal{S}_n[\alpha|\beta] \mathcal{\tilde{A}}_{n}(\beta).
	\label{eq:leftside}
\end{align}
The right-hand side of \cref{eq:leftsiderightside} becomes
\begin{align}
	\sum_{\sigma,\rho}\mathcal{A}_{n+1}\mathcal{S}_{n+1}[\sigma|\rho]\mathcal{\tilde{A}}_{n+1} =
	\sum_{\sigma,\rho}&\left[\left( \frac{S_{\textrm{YM}}^{(0)}}{\epsilon^2} + \frac{S_{\textrm{YM}}^{(1)}}{\epsilon} +S_{\textrm{YM}}^{(2)}\right) \mathcal{A}_n + \mathcal{R}_{n+1}\right] \mathcal{S}_{n+1}[\sigma,\rho] \nonumber \\
	\times &\left[\left( \frac{S_{\textrm{YM}}^{(0)}}{\epsilon^2} + \frac{S_{\textrm{YM}}^{(1)}}{\epsilon} +S_{\textrm{YM}}^{(2)}\right) \mathcal{\tilde{A}}_n + \mathcal{\tilde{R}}_{n+1}\right],
	\label{eq:rightside}
\end{align}
when we use the soft limit of the Yang-Mills amplitudes in \cref{eq:SoftFactorYM}. 
We can match the left-hand side and the right-hand side of \cref{eq:leftsiderightside} at each order in $1/\epsilon$. 
A detailed analysis of the relation at order $1/\epsilon^3$ and $1/\epsilon^2$ was performed in Ref.~\cite{Du:2014eca},
resulting in new relations for the KLT-kernel.

Focusing on $\mathcal{O}(1/\epsilon)$, we find simple relations between the universal and non-universal piece of the 
Yang-Mills amplitudes. 
A more detailed derivation can be found in \Cref{app:subsubleading}.
The non-universal pieces are defined as
\begin{align}
	\label{eq:R1def}
	R_1 &=  (-1)^{n+1} \sum_{t=1}^{n-2} \sum_{\sigma,\rho\in S_{n-3}} \frac{1}{\epsilon}
	\frac{[t,n+1]\langle t,n-1 \rangle}{\langle n+1,n-1\rangle}\,
	\mathcal{R}_{n+1}^{\epsilon\rightarrow 0}\,\mathcal{S}[\sigma|\rho]_{p_{n-1}}\,\mathcal{\tilde{A}}_n , \\
	\label{eq:R2def}
	R_2 &=  (-1)^{n+1} \sum_{t=1}^{n-2} \sum_{\sigma,\rho\in S_{n-3}} \frac{1}{\epsilon}
	\frac{[t,n+1]\langle n,t \rangle}{\langle n+1,n\rangle}\mathcal{A}_n\,\mathcal{S}[\sigma|\rho]_{p_{n-1}}\,\mathcal{\tilde{R}}_{n+1}^{\epsilon\rightarrow 0},
\end{align}
while the universal pieces come from second derivatives,
\begin{align}
	\label{eq:T1}
	T_1 &=  (-1)^n \sum_{t=1}^{n-2} \sum_{\sigma,\rho\in S_{n-3}} \frac{1}{\epsilon}
	\frac{[t,n+1]\langle n,n-1 \rangle}{\langle t,n\rangle \langle n+1,n-1 \rangle}\,\,
	\left[\frac{1}{2}\Nt{n+1}{t}^2 \left(\mathcal{A}_n\,\mathcal{S}[\sigma|\rho]_{p_{n-1}} \right)\right]
	\,\mathcal{\tilde{A}}_n, \\
	\label{eq:T2}
	T_2 &=(-1)^n \sum_{t=1}^{n-2} \sum_{\sigma,\rho\in S_{n-3}} \frac{1}{\epsilon}
\frac{[t,n+1]\langle n-1,n \rangle }{\langle t,n-1\rangle \langle n+1,n \rangle } \mathcal{A}_n \,\mathcal{S}[\sigma|\rho]_{p_{n-1}}
	\left[\frac{1}{2}\Nt{n+1}{t}^2\,\mathcal{\tilde{A}}_n\right].
\end{align}
The relation between the universal and non-universal pieces of the Yang-Mills amplitudes is
\begin{align}
\label{eq:T1T2_R1R2}
	T_1 + T_2 = R_1 + R_2.
\end{align}
\Cref{eq:T1T2_R1R2} is a new, non-trival relation, illustrating that the non-universal part of the
Yang-Mills amplitude possesses some hidden structure.
Remarkably, find that the relation simplifies into two parts,\footnote{We have explicitly verified this through
seven-points.} given by
\begin{align}
	\label{eq:T1_R1}
	T_1 &= R_1, \\
	\label{eq:T2_R2}
	T_2 &= R_2.
\end{align}
Explicitly, for e.g. \cref{eq:T2_R2}, this means that
%
\begin{align}
	&	\sum_{t=1}^{n-2} \sum_{\sigma,\rho\in S_{n-3}} \frac{1}{\epsilon}
\mathcal{A}_n\,\mathcal{S}[\sigma|\rho]_{p_{n-1}} 
	\left[ \frac{[t,n+1]\langle n,n-1 \rangle }{\langle t,n\rangle \langle n+1,n-1 \rangle } 
	\left[\frac{1}{2}\Nt{n+1}{t}^2\,\mathcal{\tilde{A}}_n\right]\right.  
+\left.\frac{[t,n+1]\langle n,t \rangle}{\langle n+1,n\rangle} \mathcal{\tilde{R}}_{n+1}^{\epsilon\rightarrow 0}\right] = 0.
\end{align}
This is a non-trivial relation for $\mathcal{R}_{n+1}$, which previously have not been discussed. 
It relates the second derivative of the universal part of a $n$-point amplitude 
to the \textit{non-universal} piece of the $(n+1)$-point amplitude. 
Taking as an example the $n+1 = 6$ NMHV amplitude.
There are usually three BCFW-diagrams contributing to the amplitude. 
One of them contains all the universal soft behavior, as in \cref{eq:BCFWsplit},
while the other two diagrams belongs to $\mathcal{R}_{n+1}$ (see \cref{eq:Rdef}). In the soft momenta limit, $\epsilon \rightarrow 0$,
we have a relation between these two diagrams and the second derivative of the 5-point amplitude $\mathcal{A}_n$.
%


\section{Effective Operators}
\label{sec:EffectiveOperators}

The inclusion of effective operators in soft theorems were first studied in Ref.~\cite{Bianchi:2014gla}. The authors considered
the operators $F^3, R^3$, and $R^2\phi$, and found that the soft theorems hold for the two first operators while
the soft graviton theorem receives a contribution from the last operator at sub-sub-leading order. 
More general operators were considered in Ref.~\cite{Elvang:2016qvq}, where 
was shown that the soft theorem for a Yang-Mills particle is corrected 
at the subleading and sub-subleading order, while gravity amplitudes get corrections at the sub-subleading order. 
The modified soft theorems take the form
\begin{align}
	\mathcal{A}_{n+1} &= \left(\frac{S^{(0)}_{\textrm{YM}}}{\epsilon^2} + \frac{S^{(1)}_{\textrm{YM}}}{\epsilon} + S^{(2)}_{\textrm{YM}}\right) \mathcal{A}_n
			+ \left(\frac{\overline S^{(1)}_{\textrm{YM}}}{\epsilon} + \overline S^{(2)}_{\textrm{YM}}\right)\mathcal{\overline{A}}_n 
			+ \mathcal{R}_{n+1}  +\mathcal{O}(\epsilon), \\
			\mathcal{M}_{n+1} &= \left(\frac{S^{(0)}_{\textrm{GR}}}{\epsilon^3} + \frac{S^{(1)}_{\textrm{GR}}}{\epsilon^2} + \frac{S^{(2)}_{\textrm{GR}}}{\epsilon}\right) \mathcal{M}_n
			+ \frac{\overline S^{(2)}_{\textrm{GR}}}{\epsilon}\mathcal{\overline{M}}_n + \mathcal{O}(\epsilon).
\end{align}
All amplitudes, including the remainder terms $\mathcal{R}_{n+1}$, can contain contributions from 
effective operators. The bar and superscript $(k)$ denote that, when corrected by effective operators, 
the particle $k$ of the $(n+1)$- and $n$-point amplitudes may be of different particle type. 
The soft theorems for the effective operator corrections are
\begin{align}
	\label{eq:softEffLO}
\overline{S}^{(1)}_{\textrm{YM}}\,\mathcal{\overline{A}}_n = - \sum_{k}  \frac{[s,k]}{\langle s,k \rangle} \mathcal{\overline{A}}^{(k)}_n,
\quad\quad
\overline{S}^{(2)}_{\textrm{GR}}\,\mathcal{\overline{M}}_n = - \sum_{k}  \frac{[s,k]^3}{\langle s,k\rangle}\mathcal{\overline{M}}^{(k)}_n ,
\end{align}
where $s$ is the soft particle and $k$ is adjacent to the soft particle. 
We have absorbed the couplings into the amplitudes.
The sum in \cref{eq:softEffLO} for a Yang-Mills particle goes over the two adjacent legs, while
for gravity it sums over all other particles.
The sub-subleading soft term for Yang-Mills particles can be found in eq.~\eqref{eq:SoftTermEff}. 

The second ingredient we need is the KLT-relation.
The open-closed string KLT-relations
are similar to the field-theory KLT-relations, with a different kernel. For instance, the 4-point string KLT-relation turns into
\begin{align}
	\label{eq:stringkernel}
	\mathcal{M}^{\textrm{closed}}_4 (1,2,3,4) = \mathcal{A}^{\textrm{open}}_4 (1,2,3,4) \left[\frac{\kappa^2}{4\pi\alpha'}\sin{(\pi x)} \right] \mathcal{\tilde{A}}^{\textrm{open}}_4 (1,2,4,3)
\end{align}
where $x= -\alpha's_{12}$ and $\alpha'$ is the inverse string tension. 
A generalized prescription for the KLT-relations for effective amplitudes was analysed in Refs.~~\cite{BjerrumBohr:2003vy,BjerrumBohr:2003af,BjerrumBohr:2004wh}. 
The new, generalized kernel used in Refs.~\cite{BjerrumBohr:2003af,BjerrumBohr:2003vy,BjerrumBohr:2004wh} was organized as a Taylor expansion in powers of $\alpha's_{12}$,
\begin{align}
\label{eq:TaylorSin}
	\frac{\sin{(\pi x)}}{\pi} \rightarrow  x(1 + c_1 x + c_2 x^2 + ...)
\end{align}
The first order in $\alpha'$ recovers the usual KLT-kernel.

To consider the KLT-relation for effective amplitudes, we need to make some assumptions. 
First, we assume that a general $n$-point
KLT-relation for effective amplitudes follows the structure found in Refs.~\cite{BjerrumBohr:2004wh,BjerrumBohr:2003af,BjerrumBohr:2003vy},
i.e. the kernel is generalized, where the leading order reproduces the original kernel, and the kernel can be expanded as a  Taylor expansion in powers of $s_{ij}/\Lambda^2$,
where $\Lambda$ is some energy scale. In string theory, $\alpha^\prime$ takes the role of $1/\Lambda^2$, 
as can be seen from \cref{eq:stringkernel}.
Second, we assume that the soft limit of the kernel is similar to eq.~\eqref{eq:SoftKernel}, 
with possibly more powers of $s_{t,n+1}$. Therefore, we assume that the soft limit of the kernel is
\begin{align}
	\label{eq:SoftKernelEff}
	\mathcal{S}[t|t]_{p_{n+1}}\mathcal{S}[\sigma|\rho]_{p_{n-1}} \rightarrow 
	\epsilon s_{t,n+1}\left(\sum_{\ell=0}^\infty c_\ell (\epsilon s_{t,n+1})^\ell \right)\, e^{\epsilon \frac{\langle n,n+1\rangle}{\langle n,t\rangle} 
	 \Nt{n+1}{t}}\,\mathcal{S}[\sigma|\rho]_{p_{n-1}},
\end{align}
where $c_0=1$. We absorb any mass scale into the unknown coefficiencts $c_\ell$, such that the mass dimension of $c_\ell$ is $-2\ell$.
 
The left-hand side of the KLT-relation in \cref{eq:leftsiderightside} now becomes
\begin{align}
	\label{eq:leftsideeffective}
	\mathcal{M}_{n+1} &= \left(\frac{S_{\textrm{GR}}^{(0)}}{\epsilon^3} + \frac{S_{\textrm{GR}}^{(1)}}{\epsilon^2} 
	+\frac{S_{\textrm{GR}}^{(2)}}{\epsilon}\right)\mathcal{M}_n + \left(\frac{\overline S^{(2)}_{\textrm{GR}}}{\epsilon}\right)\mathcal{\overline{M}}_n \nonumber \\
	&= \left(\frac{S_{\textrm{GR}}^{(0)}}{\epsilon^3} + \frac{S_{\textrm{GR}}^{(1)}}{\epsilon^2} 
	+\frac{S_{\textrm{GR}}^{(2)}}{\epsilon}\right) \sum_{\alpha,\beta}\mathcal{A}_n(\alpha)
	\mathcal{S}_n[\alpha|\beta] \mathcal{\tilde{A}}_{n}(\beta) + \left(\frac{\overline S^{(2)}_{\textrm{GR}}}{\epsilon}\right)\mathcal{\overline{M}}_n.
\end{align}
Note that the only difference between \cref{eq:leftside} and \cref{eq:leftsideeffective} is the additional term 
coming from the effective-operator extension of soft-graviton theorem.
The right-hand side of \cref{eq:leftsiderightside} is
\begin{align}
	\label{eq:rightsideeffective}
	\sum_{\sigma,\rho}&\left[\left( \frac{S_{\textrm{YM}}^{(0)}}{\epsilon^2} + \frac{S_{\textrm{YM}}^{(1)}}{\epsilon} +S_{\textrm{YM}}^{(2)}\right) \mathcal{A}_n + \left(\frac{\overline S^{(1)}_{\textrm{YM}}}{\epsilon} + \overline S^{(2)}_{\textrm{YM}}\right) \mathcal{\overline{A}}_n + \mathcal{R}_{n+1}\right] \mathcal{S}_{n+1}[\sigma,\rho] \nonumber \\
	\times &\left[\left( \frac{S_{\textrm{YM}}^{(0)}}{\epsilon^2} + \frac{S_{\textrm{YM}}^{(1)}}{\epsilon} + S_{\textrm{YM}}^{(2)}\right) \mathcal{\tilde{A}}_n + \left(\frac{\overline S^{(1)}_{\textrm{YM}}}{\epsilon} +\overline{S}^{(2)}_{\textrm{YM}}\right) \mathcal{\tilde{\overline{A}}}_n + \mathcal{\tilde{R}}_{n+1}\right].
\end{align}
By equating \cref{eq:leftsideeffective,eq:rightsideeffective} and comparing order-by-order in $1/\epsilon$, we see that the first correction from 
the effective operators appears at subleading order for the
right-hand side and at sub-subleading order for the left-hand side. 
Therefore, at order $\mathcal{O}(1/\epsilon^2)$, we find the relation
\begin{align}
	0 = \overline{U}_1 + \overline{U}_2 + U_3 + U_4,
\end{align}
where the first two terms are given by the modifications of the soft theorem for effective operators,
%
\begin{align}	
	\label{eq:U1}
	\overline{U}_1 &=  \frac{(-1)^{n+1}}{\epsilon^2} \sum_{t=1}^{n-2} \sum_{\sigma,\rho\in S_{n-3}} 
	\frac{[t,n+1]\langle n,t\rangle}{\langle n,n+1\rangle}\,
	\mathcal{A}_n \,\mathcal{S}[\sigma|\rho]_{p_{n-1}}
	\left[ \frac{[n+1,n-1]}{\langle n+1,n-1\rangle} \mathcal{\tilde{\overline{A}}}_n^{(n-1)} +
	\frac{[n+1,t]}{\langle n+1,t\rangle} \mathcal{\tilde{\overline{A}}}_n^{(t)} \right] \\\nonumber\\
	\label{eq:U2}
	\overline{U}_2 &=  \frac{(-1)^{n+1}}{\epsilon^2} \sum_{t=1}^{n-2} \sum_{\sigma,\rho\in S_{n-3}} 
	\left[\frac{[n+1,n]}{\langle n+1,n\rangle} \mathcal{\overline{A}}_n^{(n)}+
	\frac{[n+1,t]}{\langle n+1,t\rangle} \mathcal{\overline{A}}_n^{(t)}\right]
	\mathcal{S}[\sigma|\rho]_{p_{n-1}} \frac{[t,n+1]\langle t,n-1\rangle}{\langle n-1,n+1 \rangle}
	\mathcal{\tilde{A}}_n.
\end{align}
The third term comes from the expansion of the kernel, which has an unknown parameter $c_1$,
%
\begin{align}	
	U_3 &=  \frac{(-1)^n}{\epsilon^2} \sum_{t=1}^{n-2} \sum_{\sigma,\rho\in S_{n-3}} 
c_1\frac{[t,n+1]^2\la t,n\ra\la t, n-1\ra}{\la n,n+1\ra\la n+1,n-1\ra}\,
	\mathcal{A}_n\,\mathcal{S}[\sigma|\rho]_{p_{n-1}}\mathcal{\tilde{A}}_n.
	 \nonumber
\end{align}
The last term is
%
\begin{align}
	U_4 &=  \frac{(-1)^n}{\epsilon^2} \sum_{t=1}^{n-2} \sum_{\sigma,\rho\in S_{n-3}} 
	\frac{[t,n+1]\la n,n-1 \ra}{\la n-1,n+1\ra\la n,n+1\ra}\,
	\mathcal{A}_n\,\mathcal{S}[\sigma|\rho]_{p_{n-1}}
	\left[\Nt{n+1}{t}\,\mathcal{\tilde{A}}_n\right].
	 \nonumber
\end{align}
which was also found in the analysis in Ref.~\cite{Du:2014eca}, and was shown to vanish as long as the 
original kernel, defined in \cref{eq:kernelDef},
satisfies two non-trivial identities. As we have not specified the generalized kernel fully, 
we keep this term for generality.

At $\mathcal{O}(1/\epsilon)$, we proceed analogously. For the left-hand side of \cref{eq:leftsiderightside} 
we obtain the same 
terms $P_{1-6}$ in \cref{eq:P1,eq:P2,eq:P3,eq:P4,eq:P5,eq:P6} in addition to the 
gravity effective operator term.
For the right-hand side, we find the same $Q_{1-6}$ in \cref{eq:Q1,eq:Q2,eq:Q3,eq:Q4,eq:Q5,eq:Q6} 
and $R_{1,2}$ in \cref{eq:R1def,eq:R2def} as in section~\ref{sec:Subsubleading}. 
The new contributions appear when we have used $\overline S_{\textrm{YM}}^{(1)}$,
$\overline S_{\textrm{YM}}^{(2)}$, or higher terms in the expansion of the kernel. All the new contributions
are given in \Cref{app:appendix}.

After some manipulations, we obtain the following relation
\begin{align}
	\label{eq:effectiveRelation}
\frac{\overline S^{(2)}_{\textrm{GR}}}{\epsilon}\mathcal{\overline{M}}_n  + T_1 + T_2 =
\overline{Q}+ R_1  + R _2.
\end{align}
where $\overline{Q}$ is defined in the appendix. As expected, when the effective operators are 
turned off we recover \cref{eq:T1T2_R1R2}. 
Although the relation \cref{eq:effectiveRelation} is more complicated than the relation for the original amplitudes (with no effective operators), 
we can still systematically organize the contributions from the effective operators into a simple formula. 
We found here the most general relations for the effective
terms. However, further assumptions can be made on the type of effective
operators that contribute to the amplitude and the form of the kernel, which can further simplify the relations. 
We leave this for a future investigation.

\section{Conclusion}
\label{sec:Conclusion}
Using the KLT-relation and the soft limit of Yang-Mills and gravity amplitudes, we have found new, non-trivial
relations for the sub-subleading part of the Yang-Mills amplitudes. Previous analysis has only considered the 
subleading terms, and the sub-subleading part has not previously been fully discussed.
The new relations provide non-trivial constraints for the behavior of the Yang-Mills amplitudes under
the soft limit. We also studied the analogous relations when contributions from effective operators are included.

The new relations give information about the \textit{non-universal} part of the Yang-Mills amplitude.
In obtaining the relations, we went in the oppositve direction of most of uses of the KLT-formula, where we made use 
of the behavior of the gravity amplitude to extract information for the Yang-Mills amplitude.

As we have used the spinor-helicity formalism, our results are restricted to four dimensions.
Extending the analysis to arbitrary dimensions would provide insight into the generality of the result.
A natural framework for studying the relations in arbitrary dimensions is the CHY-formalism.

Recently, infinite partial soft theorems were discussed in Refs.~\cite{Hamada:2018vrw,Li:2018gnc}. Understanding the connection between
our results and the infinite partial soft theorems would be illuminating.
Also, non-linear relations for Yang-Mills amplitudes were presented in Refs.~\cite{BjerrumBohr:2010zb,Barreiro:2019ncv}.
We leave the study of the connection between these non-linear relations and the relations presented 
in this paper as a future project.


\acknowledgments

We thank N. Emil J. Bjerrum-Bohr, Poul H. Damgaard, and Matt von Hippel for useful discussions.
The work of A.H. was in part supported by the Danish National
Research Foundation (DNRF91) and the Carlsberg Foundation.
The work of R.A. was supported
by the Alexander von Humboldt Foundation, in the framework of the Sofja Kovalevskaja Award 2016, endowed by the German Federal Ministry of Education and Research and also supported by  the  Cluster  of  Excellence  ``Precision  Physics,  Fundamental
Interactions, and Structure of Matter" (PRISMA$^+$ EXC 2118/1) funded by the German Research Foundation (DFG) within the German Excellence Strategy (Project ID 39083149).

\appendix

\section{Sub-subleading Terms}
\label{app:subsubleading}

We outline the derivation of the relation for the non-universal terms at 
$\mathcal{O}(1/\epsilon)$. Recall that the color-ordered amplitudes are given by $\mathcal{A}_n \equiv
\mathcal{A}_{n}(t,\sigma,n-1,n)$ and $\tilde{\mathcal{A}}_n \equiv \mathcal{\tilde{A}}_{n+1}(n-1,\rho,n,t)$ 
and the definition of the derivative operator is $\Nt{a}{b} = \tilde{\lambda}^{\dot \alpha}_a \frac{d}{d\tilde{\lambda}^{\dot \alpha}_b}$.
The left-hand side of \cref{eq:leftsiderightside}, after applying the soft-graviton theorem and the KLT-relation,
is given in \cref{eq:leftside}. At $\mathcal{O}(1/\epsilon)$, the terms in \cref{eq:leftside} 
are given by $P_{1-6}$, where
\begin{align}
	\label{eq:P1}
	P_1 =  (-1)^n \sum_{t=1}^{n-2} \sum_{\sigma,\rho\in S_{n-3}}& \frac{1}{\epsilon}
	\frac{[t,n+1]}{\langle t,n+1\rangle} \mathcal{A}_n\,\left[\frac{1}{2}\Nt{n+1}{t}^2\,\mathcal{S}[\sigma|\rho]_{p_{n-1}}\right]
	\,\mathcal{\tilde{A}}_n, 
\end{align}
\begin{align}
	\label{eq:P2}
	P_2 =  (-1)^n \sum_{t=1}^{n-2} \sum_{\sigma,\rho\in S_{n-3}}& \frac{1}{\epsilon}
	\frac{[t,n+1]}{\langle t,n+1\rangle}  \left[\frac{1}{2}\Nt{n+1}{t}^2\, \mathcal{A}_n\right]\mathcal{S}[\sigma|\rho]_{p_{n-1}}
	\mathcal{\tilde{A}}_n,
\end{align}
\begin{align}
	\label{eq:P3}
	P_3 =  (-1)^n \sum_{t=1}^{n-2} \sum_{\sigma,\rho\in S_{n-3}}& \frac{1}{\epsilon}
\frac{[t,n+1]}{\langle t,n+1\rangle} \mathcal{A}_n\mathcal{S}[\sigma|\rho]_{p_{n-1}} 
\left[\frac{1}{2}\Nt{n+1}{t}^2\,\mathcal{\tilde{A}}_n\right] , 
\end{align}
\begin{align}
	\label{eq:P4}
	P_4 =  (-1)^n \sum_{t=1}^{n-2} \sum_{\sigma,\rho\in S_{n-3}}& \frac{1}{\epsilon}
	\frac{[t,n+1]}{\langle t,n+1\rangle}\left[\Nt{n+1}{t}\, \mathcal{A}_n\right] \left[\Nt{n+1}{t}\,\mathcal{S}[\sigma|\rho]_{p_{n-1}}\right]\mathcal{\tilde{A}}_n, 
\end{align}
\begin{align}
	\label{eq:P5}
	P_5 =  (-1)^n \sum_{t=1}^{n-2} \sum_{\sigma,\rho\in S_{n-3}} &\frac{1}{\epsilon}
	\frac{[t,n+1]}{\langle t,n+1\rangle}\left[\Nt{n+1}{t}\,\mathcal{A}_n\right] \mathcal{S}[\sigma|\rho]_{p_{n-1}}
	\left[\Nt{n+1}{t}\,\mathcal{\tilde{A}}_n\right] , 
\end{align}
\begin{align}
	\label{eq:P6}
	P_6 =  (-1)^n \sum_{t=1}^{n-2} \sum_{\sigma,\rho\in S_{n-3}}& \frac{1}{\epsilon}
	\frac{[t,n+1]}{\langle t,n+1\rangle}\mathcal{A}_n\left[\Nt{n+1}{t}\,\mathcal{S}[\sigma|\rho]_{p_{n-1}}\right]
	\left[\Nt{n+1}{t}\,\mathcal{\tilde{A}}_n\right].
\end{align}
Similarly, the right-hand side of \cref{eq:leftsiderightside} is given in \cref{eq:rightside}, consisting of
$Q_{1-6}$ and $R_{1,2}$. The non-universal terms $R_{1,2}$ are given in \cref{eq:R1def,eq:R2def}, with the residual terms being
\begin{align}
	\label{eq:Q1}
	Q_1 &=  (-1)^n \sum_{t=1}^{n-2} \sum_{\sigma,\rho\in S_{n-3}} \frac{1}{\epsilon}
	\frac{[t,n+1]\langle t,n-1\rangle\langle n,n+1\rangle}{\langle t,n+1\rangle\langle n+1,n-1\rangle\langle n,t \rangle} \mathcal{A}_n
	\left[\frac{1}{2}\Nt{n+1}{t}^2\,\mathcal{S}[\sigma|\rho]_{p_{n-1}}\right] 
	\mathcal{\tilde{A}}_n, 
\end{align}
\begin{align}
	\label{eq:Q2}
	Q_2 &=  (-1)^n \sum_{t=1}^{n-2} \sum_{\sigma,\rho\in S_{n-3}} \frac{1}{\epsilon}
	\frac{[t,n+1]\langle t,n-1 \rangle \langle n,n+1 \rangle}{\langle t,n+1\rangle \langle n+1,n-1 \rangle\langle n,t \rangle}  \left[\frac{1}{2}\Nt{n+1}{t}^2\,\mathcal{A}_n\right] \mathcal{S}[\sigma|\rho]_{p_{n-1}}  \mathcal{\tilde{A}}_n, 
\end{align}
\begin{align}
	\label{eq:Q3}
	Q_3 &=  (-1)^n \sum_{t=1}^{n-2} \sum_{\sigma,\rho\in S_{n-3}} \frac{1}{\epsilon}
\frac{[t,n+1]\langle n,t \rangle \langle n-1,n+1 \rangle}{\langle t,n+1\rangle \langle n,n+1 \rangle \langle n-1,t\rangle} \mathcal{A}_n \mathcal{S}[\sigma|\rho]_{p_{n-1}} \left[\frac{1}{2}\Nt{n+1}{t}^2\,\mathcal{\tilde{A}}_n\right] , 
\end{align}
\begin{align}
	\label{eq:Q4}
	Q_4 &=  (-1)^n \sum_{t=1}^{n-2} \sum_{\sigma,\rho\in S_{n-3}} \frac{1}{\epsilon}
	\frac{[t,n+1]\langle t,n-1\rangle\langle n, n+1\rangle}{\langle t,n+1\rangle\langle n+1,n-1 \rangle \langle n,t \rangle}\left[\Nt{n+1}{t}\, \mathcal{A}_n\right]\left[\Nt{n+1}{t}\,\mathcal{S}[\sigma|\rho]_{p_{n-1}}\right]
	\mathcal{\tilde{A}}_n, 
\end{align}
\begin{align}
	\label{eq:Q5}
	Q_5 &=  (-1)^n \sum_{t=1}^{n-2} \sum_{\sigma,\rho\in S_{n-3}} \frac{1}{\epsilon}
	\frac{[t,n+1]}{\langle t,n+1\rangle}\left[\Nt{n+1}{t},\mathcal{A}_n \right] \mathcal{S}[\sigma|\rho]_{p_{n-1}}
	\left[\Nt{n+1}{t},\mathcal{\tilde{A}}_n\right] , 
\end{align}
\begin{align}
	\label{eq:Q6}
	Q_6 &=  (-1)^n \sum_{t=1}^{n-2} \sum_{\sigma,\rho\in S_{n-3}} \frac{1}{\epsilon}
	\frac{[t,n+1]}{\langle t,n+1\rangle}\mathcal{A}_n\left[\Nt{n+1}{t},\mathcal{S}[\sigma|\rho]_{p_{n-1}}\right] 
	\left[\Nt{n+1}{t}\,\mathcal{\tilde{A}}_n\right] .
\end{align}
Note that
\begin{align}
	P_5 = Q_5, \\
	P_6 = Q_6,
\end{align}
which reduces \cref{eq:leftsiderightside} to
\begin{align}
	\label{eq:PQRv2}
	P_1 + P_2 + P_3 + P_4 = Q_1 + Q_2 + Q_3 + Q_4 + R_1 + R_2.
\end{align}
We can simplify this further by using that
\begin{align}
	T_1 = P_1 + P_2 + P_4  - Q_1 - Q_2 - Q_4 \quad\text{and}\quad
	T_2 = P_3 - Q_3,
\end{align}
which are the terms used in the main text, given in \cref{eq:T1,eq:T2}.
With these manipulations, we find new, non-trivial relations for the non-universal part
of the Yang-Mills amplitudes.
The new relations, \cref{eq:T2_R2,eq:T1_R1}, have surprisingly simple forms.

\section{Sub-subleading Terms from Effective Operator}
\label{app:appendix}

The new contributions to \cref{eq:leftsiderightside} coming from the soft theorems for effective operators involve 
$\overline S_{\textrm{YM}}^{(1)}$ and $\overline S_{\textrm{YM}}^{(2)}$. We will denote the terms by
$\overline Q_{2-6}$ as they resemble $Q_{2-6}$ in \cref{eq:Q2,eq:Q3,eq:Q4,eq:Q5,eq:Q6}. In general, 
as $\overline S_{\textrm{YM}}^{(1,2)}$ contain two different terms, we express $\overline{Q}_{2-6}$ as two terms,
e.g. $\overline Q_2 =\overline Q_2^{(t)} +\overline Q_2^{(n)}$. Note that $\overline Q_5$ is split into eight terms,
\begin{align}
	\overline{Q}_5 = \overline{Q}_5^{(n,n-1)} + \overline{Q}_5^{(n,t)} + \overline{Q}_5^{(t,n-1)} +\overline{Q}_5^{(t,t)} + \overline{Q}_5^{(L,n)} + \overline{Q}_5^{(L,t)} + \overline{Q}_5^{(R,n-1)} + \overline{Q}_5^{(R,t)}.
\end{align}
The terms are
\begin{align}
	\overline{Q}_2^{(t)} &=  \frac{(-1)^{n+1}}{\epsilon} \sum_{t=1}^{n-2} \sum_{\sigma,\rho\in S_{n-3}} 
	\frac{[t,n+1]^2\la t,n-1\ra\la n,n+1\ra}{\la n+1,n-1\ra\la n+1,t\ra \la n,t\ra}\,
	\left[\Nt{n+1}{t}\,\mathcal{\overline{A}}^{(t)}_n\right]\,
	\mathcal{S}[\sigma|\rho]_{p_{n-1}}\,
	\mathcal{\tilde{A}}_n,\nonumber 
\end{align}
\begin{align}
	\overline{Q}_2^{(n)} &=  \frac{(-1)^{n+1}}{\epsilon} \sum_{t=1}^{n-2} \sum_{\sigma,\rho\in S_{n-3}} 
	\frac{[t,n+1] \la t,n-1\ra[n+1,n]}{\la n+1,n-1\ra\la n,t\ra}\,
	\left[\Nt{n+1}{t}\,\mathcal{\overline{A}}^{(n)}_n\right]\,
	\mathcal{S}[\sigma|\rho]_{p_{n-1}}\,
	\mathcal{\tilde{A}}_n, \nonumber
\end{align}
\begin{align}
	\overline{Q}_3^{(t)} &=  \frac{(-1)^{n+1}}{\epsilon} \sum_{t=1}^{n-2} \sum_{\sigma,\rho\in S_{n-3}} 
	\frac{[t,n+1]^2\la n,t\ra\la n-1,n+1\ra}{\la n+1,n\ra\la n+1,t\ra \la n-1,t\ra}\,
	\mathcal{A}_n\,
	\mathcal{S}[\sigma|\rho]_{p_{n-1}}\,
	\left[\Nt{n+1}{t}\mathcal{\tilde{\overline{A}}}^{(t)}_n\right],\nonumber \\
	\overline{Q}_3^{(n-1)} &=  \frac{(-1)^{n+1}}{\epsilon} \sum_{t=1}^{n-2} \sum_{\sigma,\rho\in S_{n-3}} 
	\frac{[t,n+1][n+1,n-1]\la n,t\ra}{\la n+1,n\ra\la n-1,t\ra}
	\mathcal{A}_n\,
	\mathcal{S}[\sigma|\rho]_{p_{n-1}}\,
	\left[\Nt{n+1}{t}\mathcal{\tilde{\overline{A}}}^{(n-1)}_n\right]. \nonumber
\end{align}
\begin{align}	\nonumber
	\overline{Q}_4^{(n)} &=  \frac{(-1)^{n}}{\epsilon} \sum_{t=1}^{n-2} \sum_{\sigma,\rho\in S_{n-3}} 
	\frac{[t,n+1]\langle t,n-1\rangle\langle n,n+1\rangle [n+1,n]}{\la n+1,n-1\ra \la n,t\ra\la n+1,n\ra}\,
	\mathcal{\overline{A}}_n^{(n)}
	\left[\Nt{n+1}{t}\mathcal{S}[\sigma|\rho]_{p_{n-1}}\right]\,
	\mathcal{\tilde{A}}_n,
\end{align}
\begin{align}\nonumber
	\overline{Q}_4^{(t)} &=  \frac{(-1)^{n}}{\epsilon} \sum_{t=1}^{n-2} \sum_{\sigma,\rho\in S_{n-3}} 
	\frac{[t,n+1]\langle t,n-1\rangle\langle n,n+1\rangle [n+1,t]}{\la n+1,n-1\ra \la n,t\ra\la n+1,t\ra}\,
	\mathcal{\overline{A}}_n^{(t)}
	\left[\Nt{n+1}{t}\mathcal{S}[\sigma|\rho]_{p_{n-1}}\right]\,
	\mathcal{\tilde{A}}_n,
\end{align}
\begin{align}	
	\overline{Q}_5^{(n,n-1)} &=  \frac{(-1)^n}{\epsilon} \sum_{t=1}^{n-2} \sum_{\sigma,\rho\in S_{n-3}} 
	\frac{[t,n+1]\la n+1,t\ra[n+1,n][n+1,n-1]}{\la n+1,n\ra\la n+1,n-1\ra}\,
	\mathcal{\overline{A}}_n^{(n)}\,
	\mathcal{S}[\sigma|\rho]_{p_{n-1}}\,
	\mathcal{\tilde{\overline{A}}}^{(n-1)}_n,\nonumber
\end{align}
\begin{align}
	\overline{Q}_5^{(n,t)} &=  \frac{(-1)^{n+1}}{\epsilon} \sum_{t=1}^{n-2} \sum_{\sigma,\rho\in S_{n-3}} 
	\frac{[t,n+1]^2[n+1,n]}{\la n+1,n\ra}
	\mathcal{\overline{A}}_n^{(n)}\,
	\mathcal{S}[\sigma|\rho]_{p_{n-1}}\,
	\mathcal{\tilde{\overline{A}}}^{(t)}_n,\nonumber
\end{align}
\begin{align}
	\overline{Q}_5^{(t,n-1)} &=  \frac{(-1)^{n+1}}{\epsilon} \sum_{t=1}^{n-2} \sum_{\sigma,\rho\in S_{n-3}} 
	\frac{[t,n+1]^2[n+1,n-1]}{\la n+1,n-1\ra}\,
	\mathcal{\overline{A}}_n^{(n)}\,
	\mathcal{S}[\sigma|\rho]_{p_{n-1}}\,
	\mathcal{\tilde{\overline{A}}}^{(n-1)}_n,\nonumber
\end{align}
\begin{align}
	\overline{Q}_5^{(t,t)} &=  \frac{(-1)^{n+1}}{\epsilon} \sum_{t=1}^{n-2} \sum_{\sigma,\rho\in S_{n-3}} 
	\frac{[n+1,t]^3}{\la n+1,t\ra}\,
	\mathcal{\overline{A}}_n^{(t)}\,
	\mathcal{S}[\sigma|\rho]_{p_{n-1}}\,
	\mathcal{\tilde{\overline{A}}}^{(t)}_n, \nonumber
\end{align}
\begin{align}	
	\overline{Q}_5^{(L,n)} &=  \frac{(-1)^{n}}{\epsilon} \sum_{t=1}^{n-2} \sum_{\sigma,\rho\in S_{n-3}} 
	\frac{[t,n+1][n+1,n]}{\la n+1,n\ra}\,
	\mathcal{\overline{A}}_n^{(n)}\,
	\mathcal{S}[\sigma|\rho]_{p_{n-1}}\,
	\left[\Nt{n+1}{t}
	\mathcal{\tilde{A}}_n\right], \nonumber
\end{align}
\begin{align}
	\overline{Q}_5^{(L,t)} &=  \frac{(-1)^{n}}{\epsilon} \sum_{t=1}^{n-2} \sum_{\sigma,\rho\in S_{n-3}} 
	\frac{[t,n+1][n+1,t]}{\la n+1,t\ra}\,
	\mathcal{\overline{A}}_n^{(t)} \,
	\mathcal{S}[\sigma|\rho]_{p_{n-1}}\,
	\left[\Nt{n+1}{t}
	\mathcal{\tilde{A}}_n\right],\nonumber
\end{align}
\begin{align}
	\overline{Q}_5^{(R,n-1)} &=  \frac{(-1)^{n+1}}{\epsilon} \sum_{t=1}^{n-2} \sum_{\sigma,\rho\in S_{n-3}} 
	\frac{[t,n+1][n+1,n-1]}{\la n+1,n-1\ra}\,
	\left[\Nt{n+1}{t}\mathcal{A}_n\right]\,
	\mathcal{S}[\sigma|\rho]_{p_{n-1}}\,
	\mathcal{\tilde{\overline{A}}}^{(n-1)}_n,\nonumber
\end{align}
\begin{align}
	\overline{Q}_5^{(R,t)} &=  \frac{(-1)^{n+1}}{\epsilon} \sum_{t=1}^{n-2} \sum_{\sigma,\rho\in S_{n-3}} 
	\frac{[t,n+1][n+1,t]}{\la n+1,t\ra}\,
	\left[\Nt{n+1}{t}\mathcal{A}_n\right]\,
	\mathcal{S}[\sigma|\rho]_{p_{n-1}}\,
	\mathcal{\tilde{\overline{A}}}^{(t)}_n. \nonumber
\end{align}
\begin{align}
	\overline{Q}_6^{(n-1)} &=  \frac{(-1)^{n+1}}{\epsilon} \sum_{t=1}^{n-2} \sum_{\sigma,\rho\in S_{n-3}} 
	\frac{[t,n+1][n+1,n-1]}{\la n+1,n-1\ra}\,
	\mathcal{A}_n
	\left[\Nt{n+1}{t}\mathcal{S}[\sigma|\rho]_{p_{n-1}}\right]\,
	\mathcal{\tilde{\overline{A}}}^{(n-1)}_n, \nonumber
\end{align}
\begin{align}
	\overline{Q}_6^{(t)} &=  \frac{(-1)^{n+1}}{\epsilon} \sum_{t=1}^{n-2} \sum_{\sigma,\rho\in S_{n-3}} 
	\frac{[t,n+1][n+1,t]}{\la n+1,t\ra}\,
	\mathcal{A}_n 
	\left[\Nt{n+1}{t}\mathcal{S}[\sigma|\rho]_{p_{n-1}}\right]\,
	\mathcal{\tilde{\overline{A}}}^{(t)}_n.  \nonumber
\end{align}
The strategy now is to group $\overline{Q}_2^{(t)}$ and $\overline{Q}_4^{(t)}$ with $\overline{Q}_5^{(L,t)}$. 
Notice that each term
has a derivative and comes with the superscript $(t)$. We apply the Schouten identity on
the spinor brackets in the first two terms to match the last one. 
This produces a 'total' derivative called $\overline{Q}_L^{(t)}$ and an extra term, which we call $\overline{Q}_S^{(t1)}$. We do the same for others operators, finding the following rearragements
\begin{align}
&\overline{Q}_2^{(t)} + \overline{Q}_4^{(t)} + \overline{Q}_5^{(L,t)} = \overline{Q}_L^{(t)} + \overline{Q}_S^{(t1)}, \\
&\overline{Q}_2^{(n)} + \overline{Q}_4^{(n)} + \overline{Q}_5^{(L,n)} = \overline{Q}_L^{(n)} + \overline{Q}_S^{(n)}, \\
&\overline{Q}_3^{(t)} + \overline{Q}_6^{(t)} + \overline{Q}_5^{(R,t)} = \overline{Q}_R^{(t)} + \overline{Q}_S^{(t2)}, \\
&\overline{Q}_3^{(n-1)} + \overline{Q}_6^{(n-1)} + \overline{Q}_5^{(R,n-1)} = \overline{Q}_R^{(n-1)} + \overline{Q}_S^{(n-1)}, 
\end{align}
where
\begin{align}
	\overline{Q}_L^{(k)} &=  \frac{(-1)^{n+1}}{\epsilon} \sum_{t=1}^{n-2} \sum_{\sigma,\rho\in S_{n-3}} 
	(-1)\,[t,n+1]
	\frac{[n+1,k]}{\la n+1,k\ra}\,\Nt{n+1}{t}
	\left[\mathcal{\overline{A}}_n^{(k)}\,
	\mathcal{S}[\sigma|\rho]_{p_{n-1}}\,
	\mathcal{\tilde{A}}_n\right],\\
	\overline{Q}_R^{(k)} &=  \frac{(-1)^{n+1}}{\epsilon} \sum_{t=1}^{n-2} \sum_{\sigma,\rho\in S_{n-3}} 
	(-1)\,[t,n+1]
	\frac{[n+1,k]}{\la n+1,k\ra}\,\Nt{n+1}{t}
	\left[\mathcal{A}_n\,
	\mathcal{S}[\sigma|\rho]_{p_{n-1}}\,
	\mathcal{\tilde{\overline{A}}}^{(k)}_n\right].
\end{align}
The extra terms are
\begin{align}
	\overline{Q}_S^{(t1)} &=  \frac{(-1)^{n+1}}{\epsilon} \sum_{t=1}^{n-2} \sum_{\sigma,\rho\in S_{n-3}} 
	\frac{[t,n+1]^2 \la n-1,n\ra}{\la n+1,n-1\ra\la n,t\ra}\,
	\Nt{n+1}{t}\left[\mathcal{\overline{A}}^{(t)}_n\,
	\mathcal{S}[\sigma|\rho]_{p_{n-1}}\right]\,
	\mathcal{\tilde{A}}_n,
\end{align}
\begin{align}
	\overline{Q}_S^{(n)} &=  \frac{(-1)^{n}}{\epsilon} \sum_{t=1}^{n-2} \sum_{\sigma,\rho\in S_{n-3}} 
	\frac{[t,n+1][n,n-1]\la t,n+1\ra\la n,n-1\ra}{\la n+1,n-1\ra\la n,t\ra\la n+1,n\ra}\,
	\Nt{n+1}{t}\left[\mathcal{\overline{A}}^{(n)}_n\,
	\mathcal{S}[\sigma|\rho]_{p_{n-1}}\right]\,
	\mathcal{\tilde{A}}_n,
\end{align}
\begin{align}
	\overline{Q}_S^{(t2)} &=  \frac{(-1)^{n+1}}{\epsilon} \sum_{t=1}^{n-2} \sum_{\sigma,\rho\in S_{n-3}} 
	\frac{[t,n+1]^2\la n,n-1\ra}{\la n+1,n\ra\la n-1,t\ra}\,
	\mathcal{A}_n\,
	\mathcal{S}[\sigma|\rho]_{p_{n-1}}\,
	\left[\Nt{n+1}{t}\mathcal{\tilde{\overline{A}}}^{(t)}_n\right], 
\end{align}
\begin{align}
	\overline{Q}_S^{(n-1)} &=  \frac{(-1)^{n}}{\epsilon} \sum_{t=1}^{n-2} \sum_{\sigma,\rho\in S_{n-3}} 
	\hspace{-.5cm}\frac{[t,n+1][n+1,n-1]\la n,n-1\ra\la t,n+1\ra}{\la n+1,n\ra\la n-1,t\ra\la n+1,n-1\ra}\, 
	\mathcal{A}_n\,
	\mathcal{S}[\sigma|\rho]_{p_{n-1}}\,
	\left[\Nt{n+1}{t}\mathcal{\tilde{\overline{A}}}^{(n-1)}_n\right].
\end{align}
We group the terms together,
\begin{align}
	\overline{Q}_S &= \overline{Q}_S^{(n)} + \overline{Q}_S^{(t1)}+ \overline{Q}_S^{(t2)}+\overline{Q}_S^{(n-1)},\\
	\overline{Q}_{5^\prime} &= \overline{Q}_5^{(n,n-1)} + \overline{Q}_5^{(n,t)} + \overline{Q}_5^{(t,n-1)} +\overline{Q}_5^{(t,t)}, \\
	\overline{Q}_B &= \overline{Q}_L^{(t)} + \overline{Q}_L^{(n)} + \overline{Q}_R^{(t)} + \overline{Q}_R^{(n-1)}.
\end{align}
%

We also have contributions from the higher-order terms in the expansion of the kernel. We group them as
\begin{align}
	\overline{Q}_K = T^\prime_1 + T^\prime_2 + T^\prime_3 + T^\prime_4 +\overline{U}^\prime_1 +\overline{U}^\prime_2
\end{align}
The two last terms are related to $U_1$ and $U_2$ in \cref{eq:U1,eq:U2} as
\begin{align}
	\overline U_1^\prime &= \epsilon c_1 s_{t,n+1} U_1, \\
	\overline U_2^\prime &= \epsilon c_1 s_{t,n+1} U_2, 
\end{align}	
where the factor $s_{t,n+1}$ is understood to be inside the sum over $t$.

The other terms are
\begin{align}
	T_1^\prime &= \frac{(-1)^{n+1}}{\epsilon}c_1\, \sum_{t=1}^{n-2} \sum_{\sigma,\rho\in S_{n-3}} 
	\frac{[t,n+1]^2 \la n,t\ra}{\la n,n+1\ra}\, 
	\mathcal{A}_n\,
	\mathcal{S}[\sigma|\rho]_{p_{n-1}}\,
	\left[\Nt{n+1}{t}\mathcal{\tilde{A}}_n\right], 
\end{align}
\begin{align}
	T_2^\prime &= \frac{(-1)^{n+1}}{\epsilon}c_1\, \sum_{t=1}^{n-2} \sum_{\sigma,\rho\in S_{n-3}} 
	\,\frac{[t,n+1]^2 \la n-1,t\ra}{\la n-1,n+1\ra}\,
	\left[\Nt{n+1}{t}\mathcal{A}_n\right]\,
	\mathcal{S}[\sigma|\rho]_{p_{n-1}}\,
	\mathcal{\tilde{A}}_n,
\end{align}
\begin{align}
	T_3^\prime &= \frac{(-1)^{n+1}}{\epsilon}c_1\, \sum_{t=1}^{n-2} \sum_{\sigma,\rho\in S_{n-3}} 
	\,\frac{[t,n+1]^2 \la n-1,t\ra}{\la n-1,n+1\ra}\,
	\mathcal{A}_n\,
	\left[\Nt{n+1}{t}\mathcal{S}[\sigma|\rho]_{p_{n-1}}\right]\,
	\mathcal{\tilde{A}}_n, 
\end{align}
\begin{align}
	T_4^\prime &= \frac{(-1)^n}{\epsilon}c_2\, \sum_{t=1}^{n-2} \sum_{\sigma,\rho\in S_{n-3}} 
	\,\frac{[t,n+1]^3 \la n,t\ra\la t,n-1\ra \la t,n+1\ra}{\la n,n+1\ra\la n+1,n-1\ra}\,
	\left[\mathcal{A}_n\,
	\mathcal{S}[\sigma|\rho]_{p_{n-1}}\,
	\mathcal{\tilde{A}}_n\right].
\end{align}
We can rewrite $T_{1-3}^\prime$ using the Schouten identity to
\begin{align}
	T_{1}^\prime + T_2^\prime + T_3^\prime = T^\prime + T_s^\prime,
\end{align}
where
\begin{align}
	T^\prime &= \frac{(-1)^{n+1}}{\epsilon}c_1\, \sum_{t=1}^{n-2} \sum_{\sigma,\rho\in S_{n-3}} 
	\,\frac{[t,n+1]^2 \la n-1,t\ra}{\la n-1,n+1\ra}\,
	\Nt{n+1}{t}\left[\mathcal{A}_n\,
	\mathcal{S}[\sigma|\rho]_{p_{n-1}}\,
	\mathcal{\tilde{A}}_n\right],
\end{align}
\begin{align}
	T_s^\prime &= \frac{(-1)^n}{\epsilon}c_1\, \sum_{t=1}^{n-2} \sum_{\sigma,\rho\in S_{n-3}} 
	\,\frac{[t,n+1]^2 \la n,n-1\ra\la t,n+1\ra}{\la n,n+1\ra \la n+1,n-1\ra}\,
	\mathcal{A}_n\,
	\mathcal{S}[\sigma|\rho]_{p_{n-1}}\,
	\left[\Nt{n+1}{t}\mathcal{\tilde{A}}_n\right].
\end{align}
In total, we have that
\begin{align}
	\overline{Q} = \overline{Q}_{5^\prime} + \overline{Q}_B + \overline{Q}_S + \overline{Q}_K. 
\end{align}
%

\bibliographystyle{JHEP}
\bibliography{SoftTheorem}

\end{document}